\documentclass[twocolumn,aps,epsf,pra,floatfix,showpacs]{revtex4}

\usepackage{amsmath}
\usepackage{amsfonts}
\usepackage{graphicx}

\newcommand{\lp}{\left ( }
\newcommand{\rp}{\right ) }

\newcommand{\beq}{\begin{eqnarray*}}
\newcommand{\eeq}{\end{eqnarray*}}

\newcommand{\be}{\begin{eqnarray}}
\newcommand{\ee}{\end{eqnarray}}

\def\lsim{\mathrel{\rlap{\lower4pt\hbox{\hskip1pt$\sim$}}
    \raise1pt\hbox{$<$}}}                

\def\gsim{\mathrel{\rlap{\lower4pt\hbox{\hskip1pt$\sim$}}
    \raise1pt\hbox{$>$}}}                

\begin{document}

\author{Stefan K. Baur}
\affiliation{Laboratory of Atomic and Solid State Physics, Cornell
University, Ithaca, New York 14853}

\title{Stirring trapped atoms into fractional quantum Hall puddles}
\author{Kaden R.~A. Hazzard}
\affiliation{Laboratory of Atomic and Solid State Physics, Cornell
University, Ithaca, New York 14853}

\author{Erich J. Mueller}
\affiliation{Laboratory of Atomic and Solid State Physics, Cornell
University, Ithaca, New York 14853}
\

\begin{abstract}
We theoretically explore the generation of few-body analogs of fractional quantum Hall states.  We consider an array of identical few-atom clusters ($n=2,3,4$), each cluster trapped at the node of  an optical lattice.  By temporally varying the amplitude and phase of the trapping lasers, one can introduce a rotating deformation at each site.  We analyze protocols for coherently transferring ground state clusters into highly correlated states, producing theoretical fidelities (probability of reaching the target state) in excess of 99\%.
\end{abstract}


\maketitle


Cold atom experiments promise to produce unique states of matter, allowing controllable exploration of exotic physics.  For example, since rotation couples to neutral atoms in the same way that a uniform magnetic field couples to charged particles, many groups are excited about the possibility of producing analogs of fractional quantum hall states \cite{atomfqh,fqhclusters,adiabatic,homuellerclusters}.  In particular, if a two dimensional harmonically trapped gas of bosons is rotated at a frequency $\Omega$ sufficiently close to the trapping frequency $\omega$, then the ground state will have vortices bound to the atoms -- an analog of the binding of flux tubes to electrons in the fractional quantum hall effect.  The ground state will be topologically ordered and possess fractional excitations.  Technically, the difficulty with realizing this goal experimentally has been that it requires $\Omega$ to be tuned to a precision which scales as $1/n$, where $n$ is the number of particles.  Responding to this impediment, several authors \cite{fqhclusters,adiabatic,homuellerclusters} have proposed studying clusters with $n\lsim10$.  Such puddles possess many of the features of a bulk quantum hall liquid, and producing them would be a great achievement.  Here we propose and study protocols for producing strongly correlated clusters of rotating atoms.

The issue prompting this investigation 
is that in such clusters there are very few mechanisms for dissipating energy, and hence experimentally producing the ground state of a rotating cluster is nontrivial.  First, the small number of particles results in a discrete spectrum, and leaves few kinetic paths.  Second, in the strongly correlated states of interest the atoms largely avoid each other, further blocking the kinetics.  On these grounds, one should not expect to be able to cool into the ground state.  Instead we advocate a dynamical process where one coherently {\em drives} the system into the strongly correlated state through a well-planned sequence of rotating trap deformations.  This approach is based upon an analogy between the states of these atomic clusters, and the energy levels of a molecule.  By deforming the harmonic trap, and rotating the deformation, one couples the many-body states in much the same way that an oscillating electric field from a laser couples molecular states.  We consider a number of pulse sequences, finding that one can rapidly transfer atoms to a strongly correlated state with nearly unit efficiency.  
Following a proposal by Popp {\em et al.}\ \cite{adiabatic}, experimentalists at Stanford have achieved considerable success with a related procedure, where one slowly increases the rate of rotation, adiabatically transfering bosonic atoms from an initially non-rotating state, to an analog of the Laughlin state \cite{chu}.  One could also imagine implementing more sophisticated protocols such as rapid adiabatic passage \cite{stirap}.

To achieve sufficient signal to noise, any experimental attempt to study small clusters of atoms must employ an ensemble of identical systems: for example by trapping small numbers of atoms at the nodes of an optical lattice.  When formed by sufficiently intense lasers, this lattice will isolate the individual clusters, preventing any ``hopping" from one node to another.  We will not address the very interesting question of what would happen if the barriers separating the clusters were lowered.  By using filtering techniques, one can ensure that the same number of atoms sit at each node \cite{chu}.
A rotating deformation of each {\em microtrap} can be engineered through a number of techniques.  For example, if the intensity of the lattice beams forming a triangular lattice are modulated in sequence, then a rotating quadrupolar deformation is be produced.  A more versatile technique is to modulate the phases between counterpropagating lattice beams.  Changing these phases uniformly translates the lattice sites.  If one moves the lattice sites around faster than the characteristic times of atomic motion ($10^{-5}s$) but slow compared to the times for electronic excitations ($10^{-15}$s) then the atoms see a time averaged potential.  This technique, which is closely related to the time orbital potential traps pioneered at JILA \cite{top}, can produce almost arbitrary time dependent deformations of the individual traps which each of the clusters experiences \cite{chu}.  Each cluster feels the same potential.

Once created, the ensemble of clusters can be experimentally studied by a number of means.  {\em In situ} probes such as photoassociation \cite{photo} and RF spectroscopy \cite{rf} reveal details about the interparticle correlations.  In the regime of interest,  time-of-flight expansion, followed by imaging, spatially resolves the ensemble averaged {\em pre-expansion} density.  This result follows from the scaling form of the dynamics of lowest Landau level wavefunctions \cite{homuellerclusters}.

\begin{figure}[]
\includegraphics{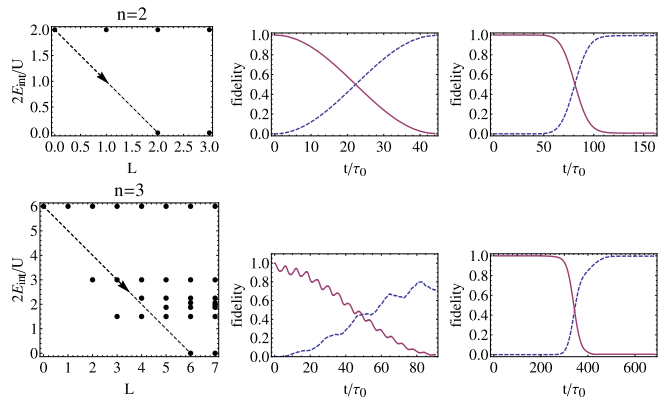}
\caption{(Color Online) Transfering small clusters from non-rotating ground state to $\nu=1/2$ Laughlin state using rotating quadrupolar ($m=2$) deformations.  Left: Interaction energy (in units of $U/2$)  of quantum states of harmonically trapped two dimensional clusters as a function of total angular momentum projection $L$ in units of $\hbar$.  Excitation paths are shown by arrows.  Central:  squared overlap (fidelity) of $|\psi(t)\rangle$ with the initial (solid) and final (dashed) states as a function of the duration of a square pulse.  Right: Fidelities as a function of time for an optimized Gaussian pulse of the form $e^{-(t-t_0)^2/\tau^2}$.  Time is measured in units of $\tau_0=\hbar/U\sim10^{-4}$s.  For $n=2$, the peak perturbation amplitude is $V_p=0.05 (U/2)$, $\omega-\Omega_p=2.0 (U/2)$, and a Gaussian pulse time of $\tau=24 \tau_0$. For $n=3$, $\tau=102 \tau_0$ and $\omega-\Omega_p=2.046 (U/2)$ and $2.055(U/2)$ for the Gaussian and square cases, respectively.  For $n=3$, nonlinear effects (coupling with near-resonant levels) shifted the optimal frequency away from the linear response expectation, $\omega-\Omega_p=2 (U/2)$. \label{fig:shaping}  }
\end{figure}

\begin{figure}[hbtp]
\includegraphics{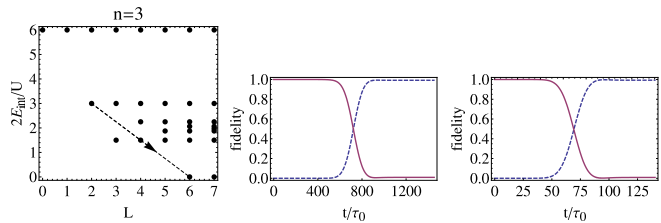}
\caption{(Color Online) Using a rotating m-fold symmetric perturbation  to drive $n=3$ particle clusters from $L=2$ to the $\nu=1/2$ Laughlin state. Left:path on the energy level diagram.  Center: second-order process coming from a deformation with $m=2$. Right: direct transition produced with $m=4$.   Solid (dashed) lines are fidelities with the initial (Laughlin) state.  In both cases the peak deformation is $V_p=0.05 (U/2)$.  Both use a Gaussian  pulse. The frequencies and pulse times $\tau$ we used for $m=2,4$  were $\omega-\Omega_p=(3.00/2) (U/2),3.035 (U/2)$ and $\tau/\tau_0=218,21$. Note how much more rapid the direct process is.  \label{fig:high-m}   }
\end{figure}

\begin{figure}[hbtp]
\includegraphics{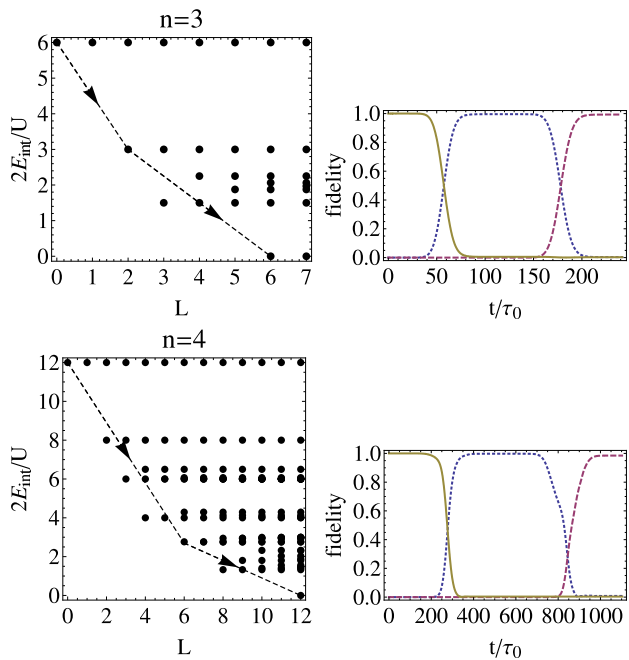}
\caption{(Color Online) Transfering atoms using multiple pulses. \label{fig:multiple}  Left: paths from initial to Laughlin states for $n=3,4$. 
Right: Solid line is the fidelity with the initial state, dotted with the intermediate $(L,E)=(2 \hbar,3 (U/2))$ state, and the dashed line with the Laughlin state.  All pulses are Gaussians. 
 Despite using multiple pulses, this technique is faster than using a higher order $m=2$ pulse.
 The frequencies ($\Omega_p$), shape ($m$), and pulse times ($\tau$) for the $N=3$ sequence were     $\hbar(\omega-\Omega_p)/(U/2)=3.00,3.035$, $m=2,4$, and $\tau/\tau_0=16.95,19.2$.  For both, $V_p=0.05 (U/2)$.  For $N=4$, using two pulses with $m=2$ and $V_p=0.2 (U/2)$, we achieve $>98\%$ fidelity after a total two-pulse sequences with $\hbar(\omega-\Omega_p)/(U/2)=3.130,1.0376$ and $\tau/\tau_0=82.5,87.0$.   
}
\end{figure}

We model a single cluster as a small number of two-dimensional harmonically trapped bosonic atoms.  The two dimensionality can be ensured by increasing the intensity of the lattice beams in the perpendicular direction.   Neglecting the zero-point energy, one finds that in the frame rotating with frequency $\Omega$, the single particle harmonic oscillator eigenstates have the form $E_{jk}=\hbar(\omega-\Omega) k+\hbar (\omega+\Omega) j;\, (j,k=0,1,\ldots)$.
In typical lattices, the interaction energy $U/\hbar\sim $10 kHz is small compared to the small oscillation frequency $\omega\sim$100 kHz \cite{jaksch}.  Therefore
%
the many-body state will be made up of single particle states with $j=0$: the lowest Landau level, with wavefunctions of the form
\be
\psi_{k}(x,y) &=& (\pi k!)^{-1/2} z^k e^{-z^* z/2},
\ee
where $z=(x+i y)/d$ with $d=\sqrt{\hbar/M\omega}$ is the complex representation of the coordinate in the plane measured in units of the oscillator length, where $M$ is the atomic mass.  Including interactions, the many-body Hamiltonian for a single cluster is then
\be\label{lll}
H_{LLL} =  \sum_{j}j \hbar (\omega-\Omega)  a^\dagger_{j} a_{j} +\sum_{jklm} V_{jklm} a^\dagger_{j}a^\dagger_{k}a_{l}a_{m}
\ee 
where $a_m$ is the annihilation operator for the single particle state $\psi_m$.  For point interactions the interaction kernel is  
\be
V_{jklm}=\frac{U}{2} \delta_{j+k-l-m} 2^{-(j+k)} \frac{(j+k)!}{\sqrt{{j!k!l!m!}}},
\ee
where $U=\sqrt{2/\pi}\,\hbar^2 a/(M d_z d^2)$ is the on-site interaction between two particles in the same well, $a$ is the three dimensional s-wave scattering length and $d_z$ is the oscillator length in the transverse direction. 
As has been explored in depth by previous authors \cite{atomfqh,fqhclusters,adiabatic,homuellerclusters}, for a given total number of particles $n$, and angular momentum projection $L$, the Hamiltonian (\ref{lll}) is a finite matrix which is readily diagonalized.  Example spectra are shown in figures~\ref{fig:shaping} through \ref{fig:multiple}.  We plot the spectra as energy versus angular momentum, with $\Omega=\omega$.  Spectra at other rotation speeds are readily found by ``tilting" the graphs -- the energy of a state with angular momentum projection $L$ is simply shifted up by $\hbar (\omega-\Omega) L$.

We imagine applying to each cluster a rotating single particle potential (in the lab frame) of the form
\begin{equation}
H_S(r,t) =  V_p(t)  \left[z^m e^{im \Omega_p t}+ (z^*)^m e^{-im\Omega_p t}\right],
\end{equation}
where $m$ determines the symmetry of the deformation (e.g. $m=2$ is a quadrupolar deformation), the envelope function $V_p(t)$ is the time-dependent amplitude of the deformation, and $\Omega_p$ is the frequency at which the perturbation rotates.    We will mainly focus on the case $m=2$.  When restricted to the lowest Landau level, this potential generates a coupling between the many-body states which in the rotating frame is expressed as
\begin{equation}
H_{S} = V_p(t) \sum_{l} v_{lm}\lp e^{i m(\Omega_p-\Omega) t}a_{l+m}^\dagger a_l +
{\rm H. C.}
\rp
\end{equation}
with $v_{lm}= 2^{-m/2} (l+m)!/\sqrt{l!(l+m)!}$. As such it only couples states whose total angular momentum projection differs by $m$.  For our calculation will work in the co-rotating frame with $\Omega=\Omega_p$, where the only time dependence is given by $V_p(t)$. 

We wish to implement a $\pi$-pulse, where the amplitude $V_p(t)$ is engineered so that after the pulse, a cluster is transfered from its initial state to a target state of our choosing.  If the perturbation coupled only two states, this would be a straightforward procedure.  The frequency $\Omega_p$ is selected so that the initial and target state are degenerate in the rotating frame.  For any finite $V_p$, the system Rabi flops between the two coupled states, and by turning off the perturbation at the right time one ends up in the target state with unit probability.  The present example is more complicated, as there are many states coupled by the perturbation.  The basic idea however remains sound:  one still chooses $\Omega_p$ to make the initial and final state degenerate.  The time dependence of $V_p(t)$ should be tailored to minimize the coupling to unwanted states.  These stray couplings could be particularly disasterous, because the coupling between the initial and target state are generically quite high order in $V_p$.
As a particularly relevant example, we consider transfering clusters from the ground state (with $L=0$) to the  the $\nu=1/2$ Laughlin state $\psi_L(z_1,\cdots z_n)=\prod_{i<j} (z_i-z_j)^2 \exp(-\sum_j |z_j|^2/4 l_B^2)$, which has angular momentum $n(n-1)$.  Using a perturbation with $m=2$, this requires a $n(n-1)/2$-order process.  
A picturesque way of thinking about the dynamics in the presence of the perturbation is to map the problem onto the motion of a particle on a complicated ``lattice".  The states of the unperturbed system are analogous to ``lattice sites", while the perturbation produces a ``hopping" between sites.  The goal is to engineer a time-dependent hopping which efficiently moves the ``particle" from a known starting position to a desired ending position.
The transfer efficiency is measured by the probability that the system is in the target state $\psi_T$ at the end of the time evolution: we plot this probability -- known as the fidelity --  as a function of time,  given by $f(t)=|\langle\psi_T|\psi(t)\rangle|^2$.

As this analogy emphasizes, the problem of
 transfering a quantum system from one state to another is generic.  M\"uller, Chiow, and Chu \cite{muller} recently considered how one can optimize pulse shapes to produce high order Bragg diffraction, while avoiding transfering atoms into unwanted momentum states.  These authors developed a formalism for calculating the fidelity by adiabatically eliminating the off-resonant states.  They found that Gaussian pulse shapes greatly outperformed simple square pulses.  This result is natural, as the smoother pulses have a much smaller bandwidth.  

We numerically solve the time dependent Schr\"odinger equation, truncating our Hilbert space at finite total angular momentum $L=n(n-1)+4$ for $n=2,3$ and $L=n(n-1)+8$ for $n=4$.  We have numerically verified that changing this cutoff to higher values has negligible effects. Figure~\ref{fig:shaping} shows $f$ in the case of $n=2$ and $n=3$ for square and Gaussian pulses.  For the square pulse the fidelity is shown as a function of pulse length.  For the Gaussian pulse, a fixed pulse duration is used, and the fidelity is shown as a function of time.  For $n=2$, where only two states are involved, the pulse shape is irrelevant.  For $n=3$, where there is a near-resonant state with $L=4$, the Gaussian pulse shape greatly outperforms the square pulse, producing nearly 100\% transfer efficiency in 10's of ms, even for a very weak perturbation.

For $n>3$ we find that these high order processes become  inefficient.  For $m=2$ the coupling between the initial and final state scale as $(V_p/U)^{n(n-1)/2}$, making transfer times unrealistical long unless one drives the system into a highly nonlinear regime.  As illustrated in figure~\ref{fig:high-m}, this difficulty can be mitigated by using perturbations with higher $m$.  There, for illustration, we consider exciting a 3-particle cluster from the lowest energy $L=2$ state to the $L=6$ Laughlin state.  The second order $m=2$ pulse requires much longer than the first order $m=4$ pulse.
An interesting aside is that one would naively have expected that the resonant $l=4$ state would make the second-order process extremely inefficient.  It turns out that the coupling to that state is fortuitously zero.

Further improved scaling can be arranged by using a sequence of 
$\pi$ pulses.  One transfers the cluster from one long-lived state to another.  Since the number of pulses scales as the angular momentum, the transfer time is then quadratic in the angular momentum, rather than exponential.  One can also tailor the path to maximize the fidelity of each step.  Some two-pulse sequences are shown in figure~\ref{fig:multiple}.  The guiding principle in designing the pulse sequences is that in each step one wants as few as possible near-resonant intermediate states.  

\textit{Summary.---}We have shown it is possible to use time dependent trap perturbations to coherently transfer boson clusters from nonrotating ground states to analogs of fractional quantum hall states. We achieve fidelity $f>99\%$ for $n=2,3$ using very weak rotating $m=2$ deformations, whose duration is of order tens of ms.  Using a two-pulse sequence, we achieve similar results for $n=4$.   We find that smooth Gaussian pulses are much more effective than square pulses, and that further efficiency can be gained by using higher order perturbations of the form $z^m$ with $m>2$.

We briefly compare our technique with Ref.~\cite{adiabatic}'s proposal. While our approaches share the use of a rotating time-averaged optical lattice potential, our proposal offers significant differences and advantages.  While Ref.~\cite{adiabatic} suggests an adiabatic evolution, we propose a coherent evolution -- analogous to a Rabi oscillation -- to the Laughlin state.  This has the advantage of being faster, easier to implement, and more robust.  For a slightly smaller perturbation relative to the adiabatic method, we achieve fidelity $\sim 1$ in contrast to the adiabatic method's $0.97$ fidelity.  Moreover, our method requires half the time.  More importantly, the adiabatic method requires carefully navigating a path through possible rotating potential strengths and frequencies as a function of time.   In contrast, our method requires only setting the pulse duration and strength, and is thus more easily implementable and less susceptible to small experimental errors.

This technique will allow the efficient creation of bosonic quantum Hall puddles -- a state of matter which has not yet been observed.  The clusters produced will be orbitally entangled, have strong interparticle correlations, have fractional excitations, and possess topological orders \cite{atomfqh,fqhclusters,adiabatic,homuellerclusters}.
Although current experiments, and the present theory, is focussed on the small-atom limit, it would be exciting to apply these techniques to larger collections of atoms, producing true analogs of fractional quantum Hall states.  The main difficulty is that the spectra become dense as $n$ increases, requiring one to set $\Omega_p$ to extremely high precision.  By carefully choosing the trajectory, taking advantage of gaps in the spectrum, one might be able to overcome such difficulties.

Finally, we mention that our approach allows one to drive the system into almost arbitrary excited states.  This may, for example, be important for using quantum hall puddles in a topological quantum computing scheme \cite{topqc}.

\textit{Acknowledgements and Author Credits:}  Stefan K. Baur and Kaden R. A. Hazzard contributed equally to this work.  We would like to thank Steven Chu and Nathan Gemelke for discussions about experiments and protocols.  This material is based upon work supported by the National Science Foundation through grant No. PHY-0758104.

\end{document}